Spin-Wave frequency division multiplexing in an yttrium iron garnet microstripe magnetized by inhomogeneous field.

Zhizhi Zhang[1,2], Michael Vogel[1], José Holanda[1], M. Benjamin Jungfleisch[3], Changjiang Liu[1], Yi Li[1,4], John E. Pearson[1], Ralu Divan[5], Wei Zhang[4], Axel Hoffmann[1,6], Yan Nie[2,*], and Valentine Novosad[1,†]

[1] Materials Science Division, Argonne National Laboratory, Argonne, IL 60439, USA
[2] School of Optical and Electronic Information, Huazhong University of Science and Technology, Wuhan 430074, China
[3] Department of Physics and Astronomy, University of Delaware, Newark, DE 19716, USA
[4] Department of Physics, Oakland University, Rochester, MI 48309, USA
[5] Center for Nanoscale Materials, Argonne National Laboratory, Argonne, IL 60439, USA
[6] Department of Materials Science and Engineering, University of Illinois at Urbana-Champaign, Urbana, IL 61801, USA

**Abstract**

Spin waves are promising candidates for information processing and transmission in a broad frequency range. In the realization of magnonic devices, the frequency depended division of the spin wave frequencies is a critical function for parallel information processing. In this work, we demonstrate a proof-of-concept spin-wave frequency division multiplexing method by magnetizing a homogenous magnetic microstripe with an inhomogeneous field. The symmetry breaking additional field is introduced by a permalloy stripe simply placed in lateral proximity to the waveguide. Spin waves with different frequencies can propagate independently, simultaneously and separately in space along the shared waveguide. This work brings new potentials for parallel information transmission and processing in magnonics.

**Introduction**

Next-generation computation concepts require parallel data processing and transmission at different frequencies simultaneously in a single, shared data-bus to achieve high efficiency and compact integration. In such systems, frequency division multiplexers (FDMs) plays an important role in the separation of multiple signals encoded in different frequencies.[1] The FDM concept is also important in the emerging field of magnonics.[2-5] In magnonic circuits, spin waves (SWs) and their quasiparticles, i.e. magnons, can encode the information in their amplitude[6,7] or phase[8,9] in a broad frequency range.[10,11] An important concept in magnonics is the logic operation, which relies on the interaction of SW based on their wave properties, especially interference.[12] It paves a way to the wave-based computation[13,14]. Interference requires the coherent

SWs to have the same or nearly the same frequency. Therefore in parallel data processing, the FDM is a crucial component in realizing practical magnonic circuits.[15]

In the context of improving the magnonic signal transmission efficiency, the ideas of SW multiplexing functions have been explored.[16-18] There, the SW beams can flow along the shared waveguides and then divide into different output channels, which can be guided by a locally-generated magnetic field[16, 17] or the global bias magnetic field along different orientations.[18] In electronics for parallel computation, FDMs enable the synchronous transmission of the signals encoded at different frequencies.[1, 15] Although this technique has been widely applied in microwave engineering and fiber optics,[19, 20] it remains to be realized in magnonic systems, despite several earlier preliminary demonstrations.[5, 21-24] In these designs, the FDM functions were enabled by the exploitation of the high anisotropy of the SWs dispersion relations.

SWs with a specific frequency in the magnetic waveguide can reach their highest intensity near the ferromagnetic resonant (FMR) field.[25-28] Similarly, the waveguide under a specific magnetic field support the SWs near the FMR frequency to reach to the highest intensity. In addition, it has been predicted[29] that a permalloy (Py, $Ni_{81}Fe_{19}$) microstripe can inhomogeneously magnetize the laterally proximate yttrium iron garnet (YIG) microstripe due to its much higher saturation magnetization ($M_s$). The edge-localized SWs in YIG microstripe can thus be tuned by such a mechanism. However, the edge-localized SWs can hardly be detected in YIG microstripe because they are spatially confined in an extremely narrow region.[30] In this work, the SW FDM function is realized in YIG magnetic microstripe magnetized under a magnetic field gradient induced by a proximate Py microstripe (see Fig. 1). The SWs carrying the information are located in the central region of the YIG microstripe, which can be detected by micro-focused Brillouin light scattering ($\mu$-BLS). We demonstrate that the SWs with different frequencies can propagate simultaneously, separately and independently at different regions in the YIG microstripe. In addition, this technique provides a noninvasive mean to engineer the SW propagation without introducing an additional interface and related damping to YIG, which is advantageous compared with the recent developments of SW manipulation with interfacial exchange.[31-36] Our results reveal a novel approach for efficient FDM applications involving shared, integrated magnonic waveguides.

**Experiments**

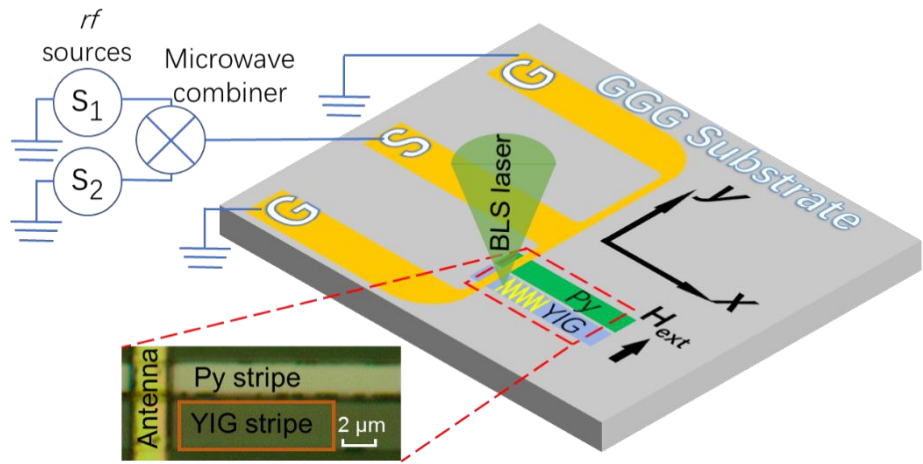

Fig. 1 Schematic illustration of the device layout and the experimental setup. The inset shows an optical microscopy image of the device indicated in the red dash box. The spin wave patterns were imaged in the region as indicated in red box.

Fig. 1 shows the schematic illustration of the device layout and the experimental setup. The 75-nm thick YIG and Py films were deposit by magnetron sputtering on single crystal gadolinium gallium garnet (GGG) substrates of 500-$\mu$m thickness with (111) orientation. The 3-$\mu$m wide YIG and 2-$\mu$m wide Py microstripes were defined by using multi-step electron-beam lithography with highly accurate alignment and fabricated by the lift-off technique. The gap between them is 200 nm. Broadband ferromagnetic resonance of the thin films yields the magnetization of saturation ($M_s$) values of 9760 G and 1960 G, and damping factors ($\alpha$) of $7.3\times10^{-3}$ and $2.1\times10^{-4}$ for Py and YIG films, respectively. For the excitation of the spin waves, the shortened end of a coplanar waveguide made of Ti(20 nm)/Au(500 nm) with a width approximately 2 $\mu$m was placed on top of the microstripes. More detailed fabricating processes were described in Ref. [25].

Two microwave generators (Anritsu MG3697C and Berkeley Nucleonics Model 845) were used to excite SWs with different frequencies simultaneously. The output signals from the two generators were combined through a microwave splitter (Anaren Model 42100). The resultant signal from the mixer was then applied to the antenna structure. The external magnetic field ($H_{ext}$) was in-plane perpendicular to the stripe and is fixed at 680 Oe, corresponding to the Damon-Eshbach modes of the SWs.[37] All the observations of the spin waves were performed using $\mu$-BLS[38] with a laser wavelength of 532 nm.

**Results and Discussions**

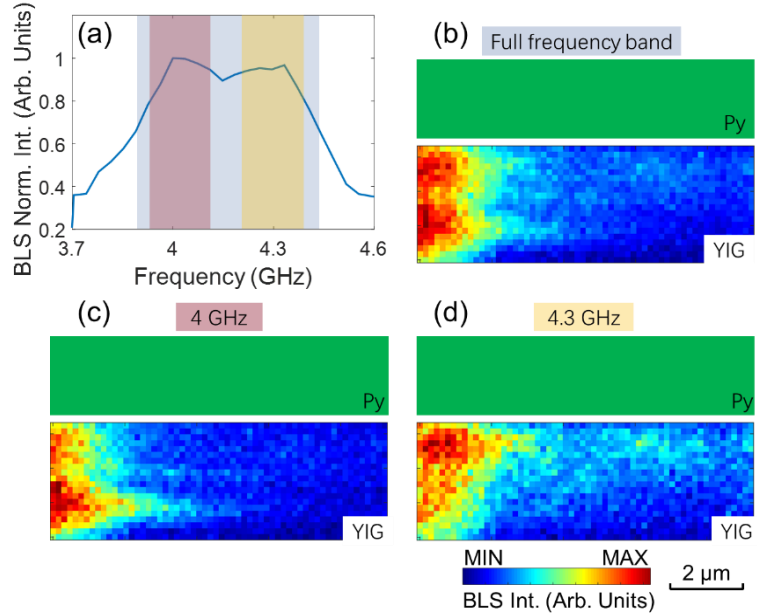

Fig. 2 Experimental demonstration of the prototype SW FDM function: (a) normalized BLS frequency spectrum recorded under the simultaneous excitations of 4 and 4.3 GHz. The spatial BLS intensity of the YIG microstripe with a proximate Py microstripe (green, same hereinafter) integrated (b) in the full

frequency band (the grey region in (a)), (c) around 4 GHz (the pink region) and (d) around 4.3 GHz (the yellow region).

We measured the SWs in the YIG microstripe under the simultaneous excitations of 4 and 4.3 GHz. Here, the frequencies of 4 and 4.3 GHz were chosen according to the dispersion relations of the Damon-Eshbach SWs, whose intensities reach the highest at the frequencies a little higher than the FMR frequency.[25] In this study, FMR frequency is $f_0 = \gamma(H_0(H_0+M_s))^{0.5} \approx 3.8$ GHz at $H_0 = 680$ Oe. The bandgap of 0.3 GHz is chosen mainly because of the limitation of the BLS frequency resolution. If the band gap is narrower, the two peaks can be hard to distinguish in BLS spectrum, considering the linewidth of the peaks. The BLS intensity spectra at every measured position are integrated and normalized as shown in Fig. 2 (a). The two peaks around 4 and 4.3 GHz indicate that the majority of the SWs in YIG microstripe are at the two frequencies. The intensity pattern of the propagating SWs integrated in the full frequency band (the grey region in Fig. 2 (a)) was mapped as shown in Fig. 2 (b). It shows that two SW beams emit from the antenna simultaneously. One is further away from the Py microstripe, while the other is closer to the Py microstripe. The intensity patterns integrated around 4 and 4.3 GHz (the pink region and the yellow region in Fig. 2 (a)) were mapped as shown in Fig. 2 (c) and 2(d), respectively. They reveal that the frequency of the SW beam farther away from (closer to) Py microstripe is 4 GHz (4.3 GHz). Neglecting the weaker intensity at the far end of the microstripe, the superposition of the two patterns in Fig. 2 (c) and 2(d) can nicely match the pattern in Fig. 2 (b). It is noticed that the beams of the two SWs are gradually separated as they propagate toward the far end. And the patterns in both Fig. 2(b) and Fig. 2(d) contain zigzag shapes.

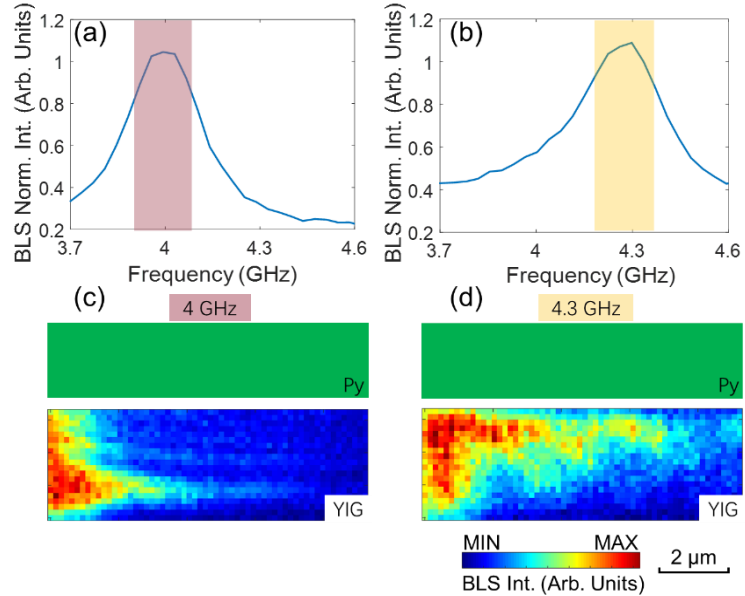

Fig. 3 Normalized BLS frequency spectrum recorded under the single excitation of (a) 4 and (b) 4.3 GHz. The pattern of the BLS intensity integrated around (c) 4 GHz (pink region in (a)) and (d) 4.3 GHz (yellow region in (b)) in the YIG microstripe under the single excitation.

To investigate the interactions between the two SWs, we measured the BLS intensity patterns with only one frequency excitation and compared them with the patterns under simultaneous excitations. The BLS intensity spectra at every measured position under a single excitation of 4 and 4.3 GHz are integrated and normalized as shown in Fig. 3(a) and 3(b), respectively. Their spatial SW profiles reproduce the results in Figs. 2(c) and 2(d), which means the two SW propagations in Fig. 2(b) do not interacting with each other. The patterns of the BLS intensity integrated around 4 GHz (pink region in Fig. 3(a)) and 4.3 GHz (yellow region in Fig. 3(b)) are shown in Fig. 3(c) and 3(d), respectively. It's noticed that the shapes of the patterns are similar with those under the simultaneous excitation (Fig. 2(c) and 2(d)). Moreover, the increase of the intensities at the far end of the patterns in Fig. 3(c) and 3(d) is due to the decrease of the unwanted tail-like signal near the antenna compared with those under the simultaneous excitation, then the contrast at the far end increases. The similar shapes of the patterns indicate that the interactions between the two spin waves, such as the interference[39] or magnon scattering,[40] are negligible, because these interactions can generate additional signals or change the SWs patterns. Despite the output power of $P = +20$ dBm, the final power reaching the sample is significantly reduced due to the insertion of the combiner. Therefore the resultant BLS pattern in Fig. 2(b) is almost the linear superposition of the two spin waves.[41]

Our results suggest that the FDM function can be realized in the YIG microstripe with a proximate Py microstripe: the two SW beams can simultaneously propagate in the YIG microstripe; their channels are spatially separated at different positions; their propagations do not interact with each other. In a previous work,[27] it has been observed that the edge localized SW beams can be shifted toward the center region of the microstripe with the increase of the frequency. This kind of shift is due to the higher $H_{eff}$ in the center of the microstripe. To get a better understanding of the FDM mechanism, we performed micromagnetic simulations to study the $H_{eff}$ using Mumax3.[42] The simulated $H_{eff}$ across the YIG microstripe versus its width at $H_{ext} = 680$ Oe is plotted in Fig. 4 (a). The position in the YIG microstripe at which the simulation is analyzed is indicated by the orange dash line in the inset of Fig. 4 (a). The presence of the Py microstripe introduces an additional static dipolar field that inhomogeneously magnetizes the YIG microstripe. In the previous study[29], the static dipolar field intensity is demonstrated to be inversely proportional to the distance. Fig. 4 (b) shows the experimentally acquired BLS intensities under different excitation frequencies in a range from 3.8 to 4.6 GHz across the YIG microstripe at the position indicated as the orange dash line in the inset of Fig. 4 (a). The $H_{eff}$ in the YIG microstripe closer to Py is dramatically increased, resulting in the higher frequency of the propagating SWs. In addition, the frequency band of the SWs closer to the Py microstripe is wider than that far away from Py microstripe. It might be attributed to a wider $H_{eff}$ range as shown in the cyan patch of Fig. 4 (a). Moreover, it has been demonstrated that the wavelength of the SWs at a specific frequency changes with the variation of the magnetic field.[43, 44] Therefore, for the SWs propagating in the region with a wide $H_{eff}$ range, they might contain multiple wavelength components. Then the zig-zag patterns of the 4.3 GHz SWs in Fig. 2 (d) and Fig. 3 (d) can be understood by the interference between the components with different wavelengths. Here, it should be noted that the 4 and 4.3 GHz SWs do not interfere with each other. While the 4.3 GHz SWs contain a set of components with different wavelengths, their coherent interference with each other lead to the stable zig-zag patterns.[25, 45, 46] In contrast, the 4 GHz SWs propagate in the region with relatively homogeneous $H_{eff}$. They have comparable single and monochromic wavelength component and appear a straight decay pattern. Furthermore, in this study, the 4 and 4.3 GHz SWs were clearly divided under $H_{ext}$ = 680 Oe. Another pair of SWs at different frequencies are also supposed to be divided if the field is tuned accordingly. Besides, the tunability can be continuous if the field is tuned continuously.

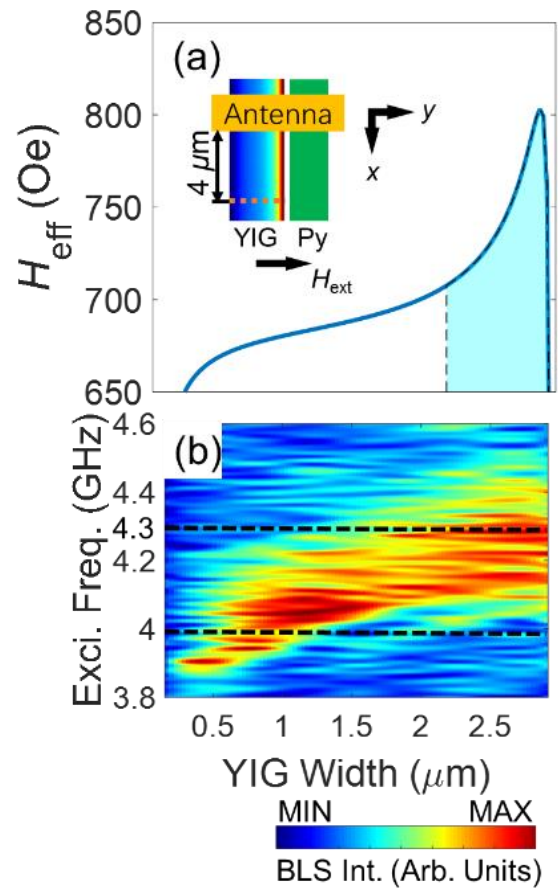

Fig. 4 (a) Simulated $H_{eff}$ across the YIG microstripe with a proximate Py microstripe under 680 Oe field. The $H_{eff}$ in the blue patched part is significantly increased due to the presence of the Py microstripe. Inset shows the schematic of the magnetic structure. The colormap encodes the y component of the $H_{eff}$ distribution inside the YIG microstripe. (b) Color coded BLS intensity at $H_{ext}$ = 680 Oe under different excitation frequencies in a range from 3.8 to 4.6 GHz across the YIG microstripe at the position indicated as the orange dash line in the inset of (a). The horizontal black dash lines indicate the 4 and 4.3 GHz excitation frequencies used in the spatial mapping.

**Conclusion**

In summary, we demonstrated that the SW FDM function can be realized in an YIG microstripe with a laterally proximate Py stripe, which introduces an inhomogeneous dipolar magnetic field on the YIG microstripe. SWs with different frequencies can propagate simultaneously, separately and independently in different channels in such magnetic microstripe. The lower (higher) frequency SWs propagate along the side farther away from (closer to) the Py microstripe. A wide field range of the $H_{\text{eff}}$ variation on the side closer to the Py microstripe results in a wider SW frequency band and multiple wavelengths for SWs at a specific frequency. The zig-zag patterns might appear due to the interference of the SWs with multiple wavelengths. These results show a new method to divide the SWs with different frequencies hybridized in a signal waveguide. The FDM function can also be continuously tunable if the field can be varied continuously. This paves a way toward the parallel processing and transmission of the SWs encoded data.

*nieyan@hust.edu.cn
†novosad@anl.gov

*Acknowledgments*
*All work was performed at the Argonne National Laboratory and supported by the Department of Energy, Office of Science, Materials Science and Engineering Division. The use of the Centre for Nanoscale Materials was supported by the US. Department of Energy (DOE), Office of Sciences, Basic Energy Sciences (BES), under Contract No. DE-AC02-06CH11357.*
*Z Z acknowledges additional financial support from the China Scholarship Council (no. 201706160146) for a research stay at Argonne. J H acknowledges financial support from the Conselho Nacional de Desenvolvimento Científico e Tecnológico(CNPq)-Brasil. W Z acknowledges support from AFOSR under Grant no. FA9550-19-1-0254. M B J acknowledges support from US National Science Foundation under Grant No. 1833000.*